\documentclass[twocolumn,trackchanges]{aastex701}

\usepackage{hyperref}

\begin{document}

%%TC:ignore
\title{A Strong Stellar Age–Metallicity Gradient Relation in Nearby Dwarf Galaxies Driven by Stellar Migration and Environmental Quenching}

\author[orcid=0000-0002-1340-7518]{Tie Li}
\affiliation{Key Laboratory for Research in Galaxies and Cosmology, Department of Astronomy, University of Science and Technology of China, Hefei 230026, China}
\affiliation{School of Astronomy and Space Science, University of Science and Technology of China, Hefei 230026, China}
\email{lt_0829@mail.ustc.edu.cn}

\author[orcid=0000-0003-1632-2541]{Hong-Xin Zhang}\thanks{Corresponding author. E-mail: hzhang18@ustc.edu.cn} 
\affiliation{Key Laboratory for Research in Galaxies and Cosmology, Department of Astronomy, University of Science and Technology of China, Hefei 230026, China}
\affiliation{School of Astronomy and Space Science, University of Science and Technology of China, Hefei 230026, China}
\email{hzhang18@ustc.edu.cn}

\author[orcid=0009-0003-8180-0455]{Wenhe Lyu}
\affiliation{Key Laboratory for Research in Galaxies and Cosmology, Department of Astronomy, University of Science and Technology of China, Hefei 230026, China}
\affiliation{School of Astronomy and Space Science, University of Science and Technology of China, Hefei 230026, China}
\email{lwh1214@mail.ustc.edu.cn}

\author[orcid=0009-0004-7885-5882]{Weibin Sun}
\affiliation{Key Laboratory for Research in Galaxies and Cosmology, Department of Astronomy, University of Science and Technology of China, Hefei 230026, China}
\affiliation{School of Astronomy and Space Science, University of Science and Technology of China, Hefei 230026, China}
\email{sunweibin@mail.ustc.edu.cn}

\author[orcid=0000-0002-8420-6246]{Bojun Tao}
\affiliation{Key Laboratory for Research in Galaxies and Cosmology, Department of Astronomy, University of Science and Technology of China, Hefei 230026, China}
\affiliation{School of Astronomy and Space Science, University of Science and Technology of China, Hefei 230026, China}
\email{bjtao@mail.ustc.edu.cn}

\author[orcid=0000-0003-0230-4596]{Weiyu Ding}
\affiliation{Key Laboratory for Research in Galaxies and Cosmology, Department of Astronomy, University of Science and Technology of China, Hefei 230026, China}
\affiliation{School of Astronomy and Space Science, University of Science and Technology of China, Hefei 230026, China}
\email{dingwy@mail.ustc.edu.cn}

\author[orcid=0000-0002-7660-2273]{Xu Kong}
\affiliation{Key Laboratory for Research in Galaxies and Cosmology, Department of Astronomy, University of Science and Technology of China, Hefei 230026, China}
\affiliation{School of Astronomy and Space Science, University of Science and Technology of China, Hefei 230026, China}
\email{xkong@ustc.edu.cn}

\author[orcid=0000-0002-4742-8800]{Guangwen Chen}
\affiliation{Instituto de Astrof\'isica de Canarias, calle Vía L\'actea s/n, E-38205 La Laguna, Tenerife, Spain}
\affiliation{Departamento de Astrof\'isica, Universidad de La Laguna, Avenida Astrof\'isico Francisco S\'anchez s/n, E-38206 La Laguna, Spain}
\email{guangwen.chen@iac.es}

\author[orcid=0000-0001-5258-1466]{Jianhui, Lian}
\affiliation{South-Western Institute for Astronomy Research, Yunnan University, Kunming, Yunnan 650504, China}
\affiliation{Yunnan Key Laboratory of Survey Science, Yunnan University, Kunming, Yunnan 650500, China}
\email{jianhui.lian@ynu.edu.cn}

\author[orcid=0000-0002-8614-6275]{Yong Shi}
\affiliation{Department of Astronomy, Westlake University, Hangzhou, Zhejiang 310030, China}
\email{shiyong@westlake.edu.cn}

\author[orcid=0000-0002-1620-0897]{Fuyan Bian}
\affiliation{European Southern Observatory, Alonso de Cordova 3107, Casilla 19001, Vitacura, Santiago 19, Chile}
\email{fbian@eso.org}

\author[orcid=0000-0001-7634-0034]{Xin Li}
\affiliation{School of Astronomy and Space Science, Nanjing University, Nanjing 210093, China}
\affiliation{Key Laboratory of Modern Astronomy and Astrophysics (Nanjing University), Ministry of Education, Nanjing 210093, China}
\email{XinLi@smail.nju.edu.cn}

\author[orcid=0000-0002-2937-6699]{Xiaoling Yu}
\affiliation{College of Physics and Electronic Engineering, Qujing Normal University, Qujing, Yunnan 655011, China}
\email{xiaoling@mail.qjnu.edu.cn}

\author[orcid=0000-0001-9018-9465]{Zhiyuan Zheng}
\affiliation{School of Astronomy and Space Science, Nanjing University, Nanjing 210093, China}
\affiliation{Key Laboratory of Modern Astronomy and Astrophysics (Nanjing University), Ministry of Education, Nanjing 210093, China}
\email{siro@smail.nju.edu.cn}

\author[orcid=0000-0003-3226-031X]{Yanmei Chen}
\affiliation{School of Astronomy and Space Science, Nanjing University, Nanjing 210093, China}
\affiliation{Key Laboratory of Modern Astronomy and Astrophysics (Nanjing University), Ministry of Education, Nanjing 210093, China}
\email{chenym@nju.edu.cn}

\author[orcid=0000-0002-3890-3729]{Qiusheng Gu}
\affiliation{School of Astronomy and Space Science, Nanjing University, Nanjing 210093, China}
\affiliation{Key Laboratory of Modern Astronomy and Astrophysics (Nanjing University), Ministry of Education, Nanjing 210093, China}
\email{qsgu@nju.edu.cn}

\author[orcid=0000-0003-4874-0369]{Junfeng Wang}
\affiliation{Department of Astronomy and Institute of Theoretical Physics and Astrophysics, Xiamen University, Xiamen, 361005, China}
\email{jfwang@xmu.edu.cn}

\author[orcid=0000-0001-8317-2788]{Shude Mao}
\affiliation{Department of Astronomy, Tsinghua University, Beijing, Beijing 100084, China}
\affiliation{ Department of Astronomy, Westlake University, Hangzhou 310030, Zhejiang Province, China}
\email{shude.mao@gmail.com}

\author[orcid=0000-0002-2583-2669]{Kai Zhu}
\affiliation{Department of Astronomy, Westlake University, Hangzhou, Zhejiang 310030, China}
\email{zhukai@westlake.edu.cn}

\begin{abstract}

Stellar metallicity gradients ($\nabla[Z/H]$) provide a fossil record of the assembly history of galaxies. We present an analysis of $\nabla[Z/H]$ for 90 nearby low-mass galaxies using VLT/MUSE IFU spectroscopy, spanning stellar masses from $10^{6.5}$ to $10^{10} M_\odot$ (median $\sim 10^{8.5} M_\odot$) and significantly extending the mass coverage of existing IFU surveys into the classical dwarf regime. Our primary finding is a robust negative correlation between $\nabla[Z/H]$ and light-weighted stellar age ($|r|\gtrsim 0.7$) measured out to $\sim$ 2$\times$ effective radius: older dwarf galaxies have steeper (more negative) gradients. This holds regardless of stellar mass, structural compactness, or large-scale environment (group/field), and is strongest in the intermediate-mass regime ($8.2\lesssim\log M_\star/M_\odot\lesssim9.0$). The slope of the age–$\nabla[Z/H]$ relation is close to that in the FIRE-2 simulations, indicating that stellar radial migration driven by feedback-induced potential fluctuations may be fundamental in dwarf evolution. But this apparent consistency is likely coincidental given the simulations' overly efficient feedback and chemical mixing. On the other hand, the H\,\textsc{i} deficiency parameter, an indicator of past environmental stripping, shows a moderate yet highly significant correlation with $\nabla[Z/H]$, second only to stellar age in strength: galaxies with higher H\,\textsc{i} deficiency tend to have more negative gradients, strongly indicating that environment-driven outside-in quenching and the ensuing gradual truncation of metal enrichment re-shape the stellar metallicity distribution. Our analysis suggests that the chemical evolution of dwarf galaxies likely arises from a synergy of feedback-driven dynamical heating and external environmental processing, though only the latter has robust observational support.
    
\end{abstract}


%%TC:endignore

\section{Introduction} 

Dwarf galaxies ($\log M_\star/M_\odot < 9.5$) occupy the lower tiers of the structural assembly hierarchy, making them crucial for understanding galaxy formation in early epochs. Their relatively small size, low metallicity, and shallow gravitational potentials make them ideal for investigating fundamental astrophysical processes, including star formation in extreme environments, stellar feedback, environmental effects and the nature of dark matter. 
These galaxies exhibit remarkable diversity in star formation histories, morphologies and environments, ranging from gas-rich, star-forming dwarf irregulars (dIrrs) in isolation to quiescent dwarf spheroidals (dSphs) in dense environments \citep[e.g.,][]{Mateo2008,McConnachie2012,Weisz2014,Simon2019}. Despite these diverse evolutionary paths, they are found to follow well-defined scaling relations, such as the mass--metallicity relation \citep[e.g.,][]{Tolstoy2009, Kirby2013, Dale2023, Zhuang2024}.

The evolutionary history of a galaxy is encoded in the spatial distribution of chemical abundances. The study of radial metallicity gradients in dwarf galaxies offers valuable insights into their evolutionary pathways. Unlike massive spirals, which typically exhibit negative age and metallicity gradients, dwarf systems at low mass end tend to exhibit a diverse range of stellar metallicity gradients, spanning positive, flat, and negative values \citep[e.g.,][]{Tully1996,Kacharov2017, Spencer2017, Taibi2020, Chen2020, Taibi2022, Li2025}, in apparent contradiction with the standard inside-out formation scenario \citep[e.g.,][]{Mo1998,Zheng2017,Sanchez2020,chen2026}.

The origin of metallicity gradients in dwarf galaxies remains actively debated, with two main classes of mechanisms under consideration. Potential internal processes include: flat metallicity gradients resulting from a homogeneous production of stellar populations in a system with high angular momentum \citep{Schroyen2011}, feedback- or turbulence-driven redistribution of ISM metals throughout the galaxies \citep[e.g., the fountain mechanism: ][]{De1994}, and radial migration of stars induced by gravitational potential fluctuations \citep[e.g.,][]{Pontzen2012,El-Badry2016,Mercado2021,Riggs2024}.

Alternatively, environmental effects can also reshape the metallicity profiles. Tidal interactions can induce bar-like structures in dwarf galaxies, promoting inward gas transport, while tidal or ram pressure stripping may effectively remove gas from the galactic outskirts \citep[e.g.,][]{Mayer2007,Buck2019, Di2021}. Environmentally induced alterations in the spatial distribution of star formation will lead to corresponding changes in the metallicity distribution.

Leveraging integral field unit (IFU) spectroscopic data from the Multi Unit Spectroscopic Explorer (MUSE; \cite{Bacon2010}) instrument on the Very Large Telescope (VLT), we explored gas-phase metallicity gradients of a representative sample of 55 nearby classical dwarf galaxies and found a significant correlation with galaxy stellar mass and outflow-driven metal loss \citep{Li2025}. In this work, we focus on the stellar metallicity gradients of an enlarged sample, aiming to understand physical drivers behind the gradients of dwarf galaxies.

This paper is organized as follows. Section \ref{sec:Data reduction} describes the sample selection and methodology for extracting stellar population gradients. We present our main observational results in Section \ref{sec:age}, followed by a detailed discussion of the underlying physical mechanisms in Sections \ref{sec:migration} and \ref{sec:origin}. Our findings are summarized in Section \ref{sec:summary}. Throughout this paper, $[Z/H]$ is used interchangeably with ``metallicity'', and metallicity gradients refer to the stellar component unless otherwise specified. We assume a standard $\Lambda$CDM cosmology ($H_{0}=70 \mathrm{~km} \mathrm{~s}^{-1} \mathrm{Mpc}^{-1}, \Omega_{\mathrm{m}}=0.3$, $\Omega_{\Lambda}=0.7$.)

\section{Sample Selection and Reduction}
\label{sec:Data reduction}

We utilize a sample of dwarf galaxies observed with VLT/MUSE in Wide-Field Mode ($1' \times 1'$ FOV, $0.2''$ sampling). Comprehensive details regarding sample selection, data reduction and the measurement of various properties, such as galaxy stellar mass mass, effective radius and star formation rate (SFR), are provided in \cite{Li2025}; here we summarize the key elements.

\subsection{The MUSE Sample and Data Reduction}

Our parent sample of 120 nearby dwarf galaxies ($D < 30$ Mpc; $M_B > -18.5$ mag) combines archival ESO observations\footnote{\url{http://archive.eso.org/wdb/wdb/eso/muse/form}} with a subset from the Dwarf Galaxy Integral-field Survey \citep[DGIS;][]{Li2026}. Data were reduced using the MUSE pipeline within the \texttt{EsoReflex} \citep{Freudling2013}. Observations affected by poor weather conditions or instrumental failures were excluded during the calibration and combination stages. Our final sample includes 90 galaxies. We utilized the Zurich Atmosphere Purge package \citep[ZAP;][]{Soto2016} for residual sky subtraction and refined flux calibration by anchoring MUSE data to DESI Legacy Imaging Surveys\footnote{\url{https://www.legacysurvey.org/}} photometry.

\subsection{Full Spectral Fitting}
\label{sec:ppxf}
Stellar population properties are extracted using the Penalized PiXel-Fitting (\texttt{pPXF}; \cite{Cappellari2017}), a robust tool for extracting stellar kinematics and population properties by fitting weighted combinations of single stellar population (SSP) models. In this work, we employ the E-MILES SSP models \citep{Vazdekis2016}, which comprise 53 ages (0.03--14\,Gyr) and 12 metallicities ($[M/H]$ from $-2.27$ to $+0.40$). To ensure consistency and avoid contamination from residual sky lines in the redder wavelengths, the fitting is performed over the range 4800--7200\,\AA, consistent with the methodology in \cite{Li2025}. Using the residuals between the observed and best‑fit spectra to randomly perturb each observed spectrum for 300 times, we perform the pPXF fitting for all the 300 realizations and obtain parameter uncertainties from the resulting distributions.

We derive light- and mass-weighted average stellar ages and metallicities based on the pPXF fitting results. To ensure robust constraints on the stellar age and metallicity, we pay particular attention to the fitting quality of key age- and metallicity-sensitive absorption features, such as the Mg \textit{b}, Fe5270, and Balmer lines. Additionally, for nearly all spectra, the 300 random realizations show no correlation between the best-fit age and metallicity, confirming that the fitting results are not subject to significant age–metallicity degeneracy.

\subsection{Radial Stellar Population Gradients}
\label{sec:gra_measure}

\begin{figure*}[t]
    \centering
    \includegraphics[width=\textwidth]{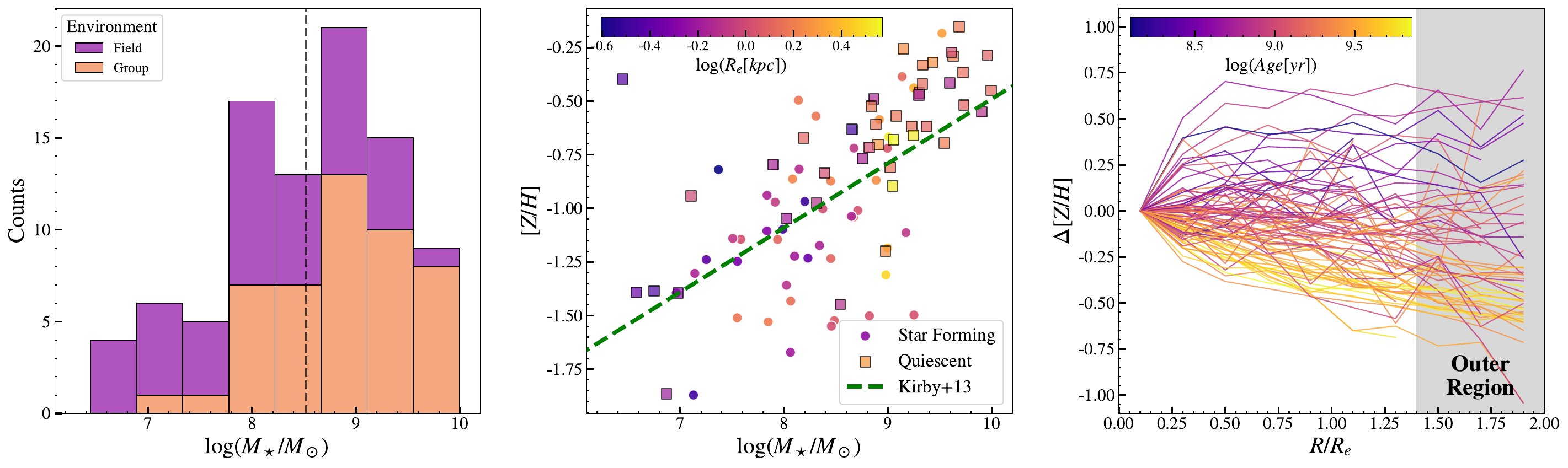}
    \caption{\textbf{Global and structural properties of the galaxy sample.} 
    \textit{Left:} Distribution of stellar masses for Filed (purple) and Group (orange) environments. The sample covers a mass range of $10^{6.5}\text{--}10^{10}\,M_\odot$. The vertical dashed line marks the median stellar mass of the entire sample ($\log(M_\star/M_\odot) \approx 8.5$). 
    \textit{Middle:} Stellar mass--metallicity relation (MZR), where each point is color-coded by the effective radius. Circular and square symbols represent star forming and quiescent galaxies, respectively. The green dashed line represents the Local Group dwarf galaxy relation from \citet{Kirby2013}.
    \textit{Right:} Radial profiles of residual light-weighted metallicity ($\Delta \text{[Z/H]}$) as a function of $R_e$-normalized major-axis radius. Lines are color-coded by the light-weighted stellar age of the galaxy. The shaded gray region indicates the outer parts of the galaxies ($R > 1.4\,R_e$). A clear trend is visible where older galaxies (orange/yellow) exhibit steeper negative gradients compared to the flatter or slightly positive profiles of younger systems (purple).}
    \label{fig:distribution}
\end{figure*}

Reliable stellar ages and metallicities require high continuum signal-to-noise ratios (SNR), a major challenge for low-surface-brightness dwarf galaxies. To overcome this, we extract integrated spectra from concentric elliptical annuli with radial widths of $0.2\,R_e$, following the isophotal geometry derived from broadband imaging.  We then determined the stellar ages and metallicities within each annulus via \texttt{pPXF}.

To ensure adequate spectral quality and spatial coverage, we require MUSE observations to reach at least $1\,R_e$ with a continuum SNR $>10$ per annulus (measured near $5500\,\text{\AA}$) of each galaxy. These requirements result in a final sample of 90 galaxies (see Table \ref{tab:property} for their properties). For this subsample, the MUSE observations reach maximum radii ranging from $1\text{--}2\,R_e$, with the vast majority ($\sim$ 90\%) extending beyond $1.4\,R_e$. To maintain a consistent baseline, we derived all radial gradients using a fixed radial range up to a maximum of $1.4\,R_e$ for all galaxies.

The gradient of a light- or mass-weighted parameter $\phi$ (e.g., $[Z/H]$ or $\mathrm{Age}$) is calculated via a linear least-squares fit to its radial profile, defined as:
\begin{equation}
\nabla \phi = \frac{\mathrm{d}\phi}{\mathrm{d}(R / R_e)}
\end{equation}
where $R/R_e$ represents the radius normalized by the effective radius. To capture the overall radial trend and avoid bias from any single annular bin, we randomly resample the annular data points with replacement 1000 times for gradient fitting, and adopt the median and central 68\% of the resulting gradient distribution as the most likely gradient and its confidence interval. The outshining effect of younger stellar populations in integrated spectra generally hampers robust constraints on the ancient, typically mass-dominant populations. Therefore, our analysis throughout this paper primarily relies on light-weighted parameters.

Figure \ref{fig:distribution} illustrates the stellar masses, global metallicities, and radial metallicity profiles of our sample.
The left panel shows the stellar mass distribution, spanning $10^{6.5}$ to $10^{10} M_\odot$. The sample is categorized by environment following \citet{Kourkchi2017}. Galaxies identified as members of a group or cluster (richness $N \ge 2$) are categorized as group galaxies ($N=47$), while ``singles'' are classified as field galaxies ($N=43$). The median stellar mass of entire sample ($\log M_\star/M_\odot \sim 8.5$) is marked as a reference.

The middle panel displays the stellar mass–metallicity relation (MZR), color-coded by $R_e$. The circles and squares represent star-forming and quiescent galaxies, with the division defined by a specific SFR threshold (sSFR) of $\log (\mathrm{sSFR/yr^{-1}}) = -11$. Our sample broadly follows the established relation for Local Group dwarfs from \cite{Kirby2013} at $\log M_\star/M_\odot<8.5$, but lies systematically above it at higher masses, consistent with \citet{Zhuang2024}. The right panel presents the radial metallicity profiles as a function of $R_e$-normalized radius, color-coded by stellar age of each galaxy. Notably, a clear trend is observed that galaxies with steeper negative metallicity profiles tend to have older stellar ages.

\subsection{Quantification of the Galaxy Environment}
Beyond the two-fold classification of group and field environments, we employ two quantitative parameters to more precisely quantify environmental processing: the local galaxy number density ($\eta_{k}$) and $\text{H}\,\textsc{i}$ deficiency ($Def_{\text{HI}}$).

Following \citet{Argudo2015}, we quantify the local Large Scale Structure using the projected galaxy number density $\eta_{k}$, defined as:

\begin{equation}
\label{eq:pden}
\eta_{k} \equiv \log \left(\frac{k-1}{\mathrm{Vol}(d_{k})}\right) = \log \left(\frac{3(k-1)}{4 \pi d_{k}^{3}}\right)
\end{equation}

where $d_{k}$ is the projected physical distance to the $k$-th nearest neighbor, based on the dataset provided by \citet{Ohlson2024}. Here, we adopt 5 as the value of $k$. Higher $\eta_{k}$ values represent denser regions where tidal interactions are more likely.

To assess the impact of environmental processing on gas content, we calculate the $\text{H}\,\textsc{i}$ deficiency parameter, $Def_{\text{HI}}$. Following the classical definition \citep[e.g.,][]{Haynes1984}, $Def_{\text{HI}}$ is defined as the logarithmic difference between the expected $\text{H}\,\textsc{i}$ mass of an isolated galaxy and its currently observed $\text{H}\,\textsc{i}$ mass:

\begin{equation}
\label{eq:def_hi}
    Def_{\text{HI}} = \log_{10} \left( M_{\text{HI, exp}} \right) - \log_{10} \left( M_{\text{HI, obs}} \right),
\end{equation}

The expected $\text{H}\,\textsc{i}$ mass is estimated based on the optical size and Hubble type, following the scaling relations calibrated by \citet{Haynes1984}:

\begin{equation}
\log(M_{\text{HI,exp}}/M_\odot) = a + 2b \log(D_{25}/\text{kpc}) - 2\log h,
\end{equation}

where $D_{25}$ is the $B$-band 25th-mag\,arcsec$^{-2}$ isophotal diameter retrieved from HyperLeda\footnote{\url{http://atlas.obs-hp.fr/hyperleda/}}, and $h = H_0 / (100\,\text{km\,s}^{-1})$. We adopt the coefficients $a$ and $b$ for each morphological type as summarized in \cite{Boselli2009}.

Observed $\text{H}\,\textsc{i}$ masses are compiled from, in order of preference, the ALFALFA catalog \citep[][]{haynes2018}, HIPASS \citep{meyer2004}, the All Digital HI Catalog in the Extragalactic Distance Database \citep{Courtois2009}, and other archival sources \citep[][]{loni2021}. This compilation yields $\text{H}\,\textsc{i}$ detections for 62 galaxies in our sample.

For the 28 non-detections, we estimate $3\sigma$ upper limits for their $\text{H}\,\textsc{i}$ masses based on the ALFALFA survey sensitivity. Following \citet{haynes2011}, the upper limit is calculated as:

\begin{equation}
    M_{\text{HI, lim}} = 2.36 \times 10^5 \cdot d^2 \cdot S_{\text{lim}},
\end{equation}

where $d$ is the distance in Mpc and $S_{\text{lim}}$ is the integrated flux limit. The latter is derived by considering the typical ALFALFA noise level and an assumed velocity width of 100 km/s. For these sources, the resulting $Def_{\text{HI}}$ values are treated as lower limits, providing a conservative estimate of the gas deficit.

\section{Discovery of a strong correlation between average stellar age and metallicity gradient}
\label{sec:age}

\begin{figure*}[t]
  \centering
  \includegraphics[width=\textwidth]{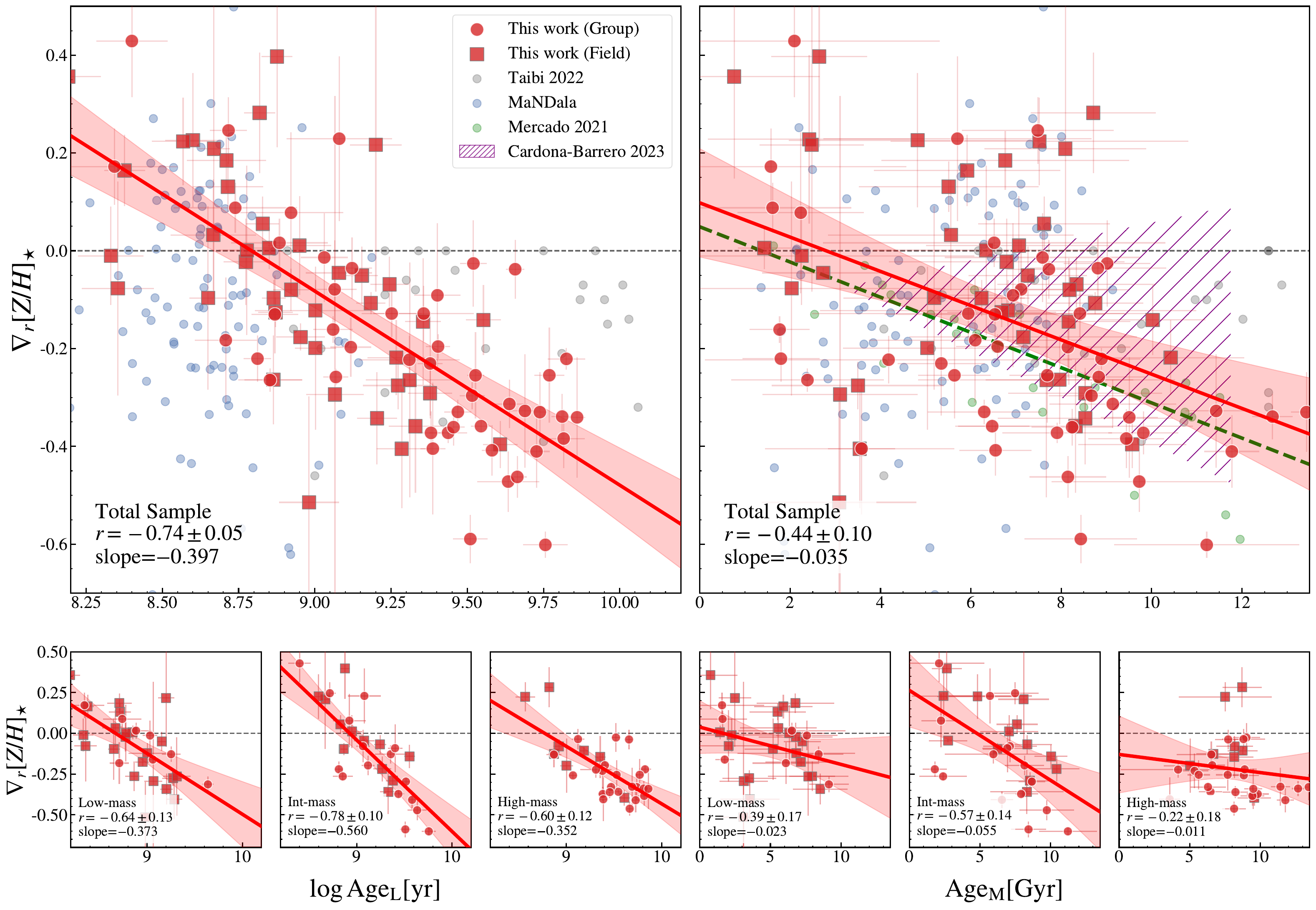}
  \caption{\textbf{Correlation between metallicity gradients ($\nabla_r[Z/H]$) and average stellar ages.} 
  The top row shows the total sample for light-weighted ages ($\log \mathrm{Age}_{\mathrm{L}} [\mathrm{yr}]$, left) and mass-weighted ages ($\mathrm{Age}_{\mathrm{M}} [\mathrm{Gyr}]$, right). 
  Circles and squares represent group and field galaxies, respectively. The solid red lines and shaded regions denote the best-fit linear relations and their 1$\sigma$ uncertainties. 
  For comparison, we overplot Local Group dwarfs from \cite{Taibi2022} (grey circles; using $t_{90}$ and $t_{50}$ as proxies for $\mathrm{Age}_{\mathrm{L}}$ and $\mathrm{Age}_{\mathrm{M}}$, respectively), the MaNDala sample \citep{Cano2025} (blue circles), and simulation results from \cite{Mercado2021} (green dashed line in the mass-weighted panel) and \cite{CardonaBarrero2023} (purple shaded region in the mass-weighted panel).
  The bottom panels show the same correlations subdivided into three stellar mass bins (low-, intermediate-, and high-mass, as defined in the text). 
  Pearson correlation coefficients ($r$) and best-fit slopes are indicated in each panel. The horizontal dashed lines mark a flat gradient ($\nabla_r = 0$). Note that $\mathrm{Age}_{\mathrm{L}}$ is plotted in logarithmic scale while $\mathrm{Age}_{\mathrm{M}}$ is plotted in linear scale (for a direct comparison with \cite{Mercado2021}).}
  \label{fig:azgr}
\end{figure*}

After an extensive exploration of the correlations between metallicity gradients and various galaxy properties, we find that stellar age shows the tightest correlation with the observed gradient diversity. In this section, we present this relation and discuss its significance across different age and stellar mass regimes.

Figure \ref{fig:azgr} presents the light-weighted metallicity gradients ($\nabla_r [Z/H]$) as a function of light-weighted ages ($\mathrm{Age}_{\mathrm{L}}$, left panels) and mass-weighted ages ($\mathrm{Age}_{\mathrm{M}}$, right panels). For comparison, we include results from 30 Local Group dwarf galaxies \citep{Taibi2022} and the MaNGA Dwarf Galaxy Survey \citep[MaNDala;][]{Cano2025}). For the Local Group sample, we adopt $t_{90}$ and $t_{50}$ (the lookback times to reach 90\% and 50\% mass assembly) as the observational proxies for $\mathrm{Age}_{\mathrm{L}}$ and $\mathrm{Age}_{\mathrm{M}}$, respectively. Notably, the MaNDala sample is biased to massive systems, with only 7 galaxies below the median mass of our sample.

As shown in the main panels of Figure \ref{fig:azgr}, a robust negative correlation exists across both age definitions: older galaxies exhibit more negative gradients, while younger systems show flatter or slightly positive profiles. For $\mathrm{Age}_{\mathrm{L}}$, we find a strong correlation coefficient of $r = -0.74$ ($p \approx 10^{-15}$) with a slope of $-0.387$. The correlation remains significant but shallower for $\mathrm{Age}_{\mathrm{M}}$ ($r = -0.44,\, p \approx 10^{-5}$). The increased scatter and weaker correlation for $\mathrm{Age}_{\mathrm{M}}$ likely stem from the inherent uncertainties and systematic biases associated with mass-weighted estimates in integrated-light spectral modeling, particularly for older systems where the age–metallicity degeneracy is most pronounced.

To assess the robustness of this age-metallicity gradient correlation against the choice of radial ranges for gradient fitting, we perform the gradient fitting by adopting smaller or larger outermost radii than the fiducial choice of $1.4 R_{\mathrm{e}}$, from $1.2R_{\mathrm{e}}$ to $1.8R_{\mathrm{e}}$, with 62 galaxies reaching $1.8R_{\mathrm{e}}$. We find a subtle trend with increasing outer radial boundary, whereby the $\log\mathrm{Age}_\mathrm{L}$--$\nabla_r[Z/H]_\star$ correlation coefficient varies from $r = -0.78 \pm 0.04$ to $-0.68 \pm 0.06$ and the slope shallows from $-0.466$ to $-0.315$, and the correlation with the mass-weighted ages shows a similar trend ($r$: $-0.44$ -- $-0.34$; slope: $-0.064$ -- $-0.043$; Figure \ref{fig:differ_r_appendix}). This indicates that the radial metallicity profiles become flatter at larger radii, and that a linear gradient offers only a first-order description of the radial variation. With this caveat in mind, the systematic dependence of the metallicity profiles on average stellar ages—particularly the light-weighted ones—remains significant and robust, and our conclusions throughout this paper remain qualitatively unchanged.

To examine the potential mass dependence of stellar age--metallicity gradient relation, we divide the sample into three bins ($N=30$ each): low-mass ($\log M_\star/M_\odot \leq 8.17$), intermediate-mass ($8.17 < \log M_\star/M_\odot < 9$), and high-mass ($\log M_\star/M_\odot \geq 9$). As shown in the bottom subpanels of Figure \ref{fig:azgr}, a significant negative correlation persists across nearly all mass regimes, with the exception of the $\mathrm{Age}_{\mathrm{M}}$ relation in the high-mass subsample. Notably, the intermediate-mass subsample exhibits the tightest correlation and the steepest slopes for both age definitions, while the high-mass subsample shows the weakest trend.

The stellar age–$\nabla_r [Z/H]$ relation shows consistent correlation strength and slope between field ($r = -0.72$, slope = $-0.40$) and group ($r = -0.68$, slope = $-0.39$) subsamples, without significant difference, suggesting the underlying driver is independent of large-scale environment.

\section{The Role of Stellar Radial Migration in Shaping Metallicity Gradients}
\label{sec:migration}

In Figure \ref{fig:azgr}, we compare the observed mass-weighted age--$\nabla_r [Z/H]$ with predictions from the FIRE-2 simulations by \cite{Mercado2021}. This comparison is restricted to $\mathrm{Age}_{\mathrm{M}}$ as \cite{Mercado2021} only report $t_{50}$. The metallicity gradient was fitted over a radial range extending to 2$R_{\rm e}$ in the simulation. The observed slope ($\sim$ $-0.035$ to $-0.047$ for gradient fits with outermost radii from 1.4$R_{\rm e}$ to 1.8$R_{\rm e}$) is close to that in the simulation ($-$0.036); the observed intercept is slightly smaller than in the simulation.

The broadly similar age-dependence of the metallicity gradient to the FIRE-2 simulation suggests that stellar radial migration may be an important driver in the structural and chemical evolution of dwarf galaxies. According to the framework proposed by \citet{Mercado2021} and \citet{El-Badry2016}, this migration is resulted from the ``bursty'' nature of star formation histories, which are particularly prevalent in low-mass galaxies at early epochs \citep[e.g.,][]{Sun2023,Endsley2025,Perry2025,Clarke2025}. These episodic, intense star formation events trigger powerful, rhythmic feedback-driven outflows, inducing temporal fluctuations in the central gravitational potential, effectively ``heating'' the stellar orbits and driving pre-existing stars toward larger radii. Conversely, late-time gas accretion—characterized by higher angular momentum and pre-enrichment in the circumgalactic medium—promotes spatially extended star formation, which acts to flatten the overall gradient.

This mechanism may naturally explain the mass dependence of the observed correlation. In the lowest‑mass systems, a combination of cuspy dark matter halo profiles \citep[e.g.,][]{Governato2012, Lazar2020} and relatively inefficient star formation limits the ability of stellar feedback to drive significant potential fluctuations. At the high galaxy mass end, star formation proceeds steadily, and the deep gravitational potential wells constrain  fluctuations, thereby suppressing significant stellar redistribution.The intermediate-mass regime is the ``sweet spot'' where repeated feedback episodes produce strong potential fluctuations, driving the formation of cored dark matter profiles \citep[e.g.,][]{Read2016} and efficient stellar redistribution.

However, several caveats suggest that this radial migration scenario, while playing a role, may not be the sole driver. The FIRE-2 simulations predict a well-mixed interstellar medium (ISM) across all cosmic epochs for galaxies with $M_\star < 10^9\,M_\odot$, typically resulting in nearly flat gas-phase metallicity gradients \citep{Escala2018}. In such a strong feedback framework, stellar radial migration is the only available mechanism to create non-flat stellar metallicity gradients. In other words, the stellar migration effect reaches its maximum in the FIRE-2 simulations. However, contrary to FIRE-2's mass-independent flat gas metallicity gradients, recent observations by \citet{Li2025} reveal a significant negative correlation between gas metallicity gradients and stellar mass in nearby dwarfs, particularly when accounting for effective metal yields. The radial migration scenario also struggles to account for the positive metallicity gradients observed in the youngest galaxies of our sample. These discrepancies suggest that the apparent consistency of our observed stellar age-metallicity gradient relation with \cite{Mercado2021} may not imply a unique validation of full-scale feedback-driven migration, given the many other factors involved in the structural evolution of dwarf galaxies (see below).

Radial migration may not be as important as predicted by FIRE-2 simulations. \citet{Pilkington2012} showed that variations in feedback prescriptions, metal diffusion, and other subgrid physics can substantially alter the resulting chemical distributions in simulations. Indeed, earlier simulations by \citet{Schroyen2013} found limited stellar migration, with a negligible impact on metallicity profiles in dwarf systems.  Even in more massive galaxies, where secular migration via resonant orbital scattering is expected to be most efficient, \citet{Sanchez2014} found no significant observational evidence for migration when comparing barred and unbarred systems. Lastly, we must emphasize that the mass-weighted stellar population properties derived from integrated spectra are subject to substantial uncertainties and should therefore be interpreted with caution.

\section{Alternative explanations of the stellar age-metallicity gradient relation}
\label{sec:origin}

\begin{figure*}
  \centering
  \includegraphics[width=1\textwidth]{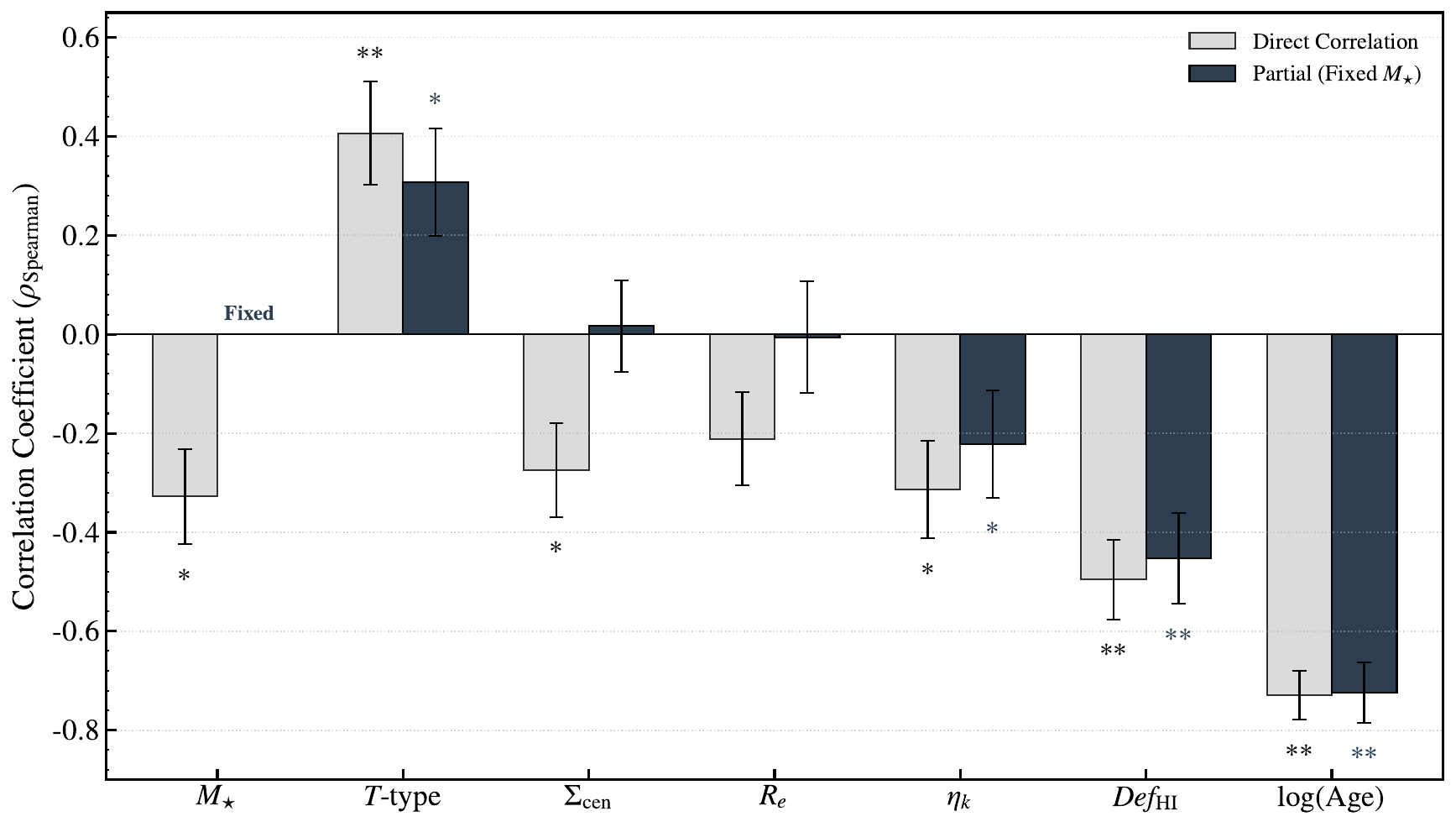}
  \caption{\textbf{Spearman rank correlation coefficients ($\rho_{\mathrm{Spearman}}$) between metallicity gradients ($\nabla_r [Z/H]$) and various galaxy properties.} 
  The light grey bars represent the direct (zero-order) correlation coefficients, while the dark blue bars indicate the partial correlation coefficients after controlling for galaxy stellar mass ($M_\star$). 
  The analyzed properties include stellar mass ($M_\star$), H\,\textsc{i} deficiency parameter ($Def_\mathrm{HI}$), central stellar mass surface density ($\Sigma_{\mathrm{cen}}$), morphological T-type, local projected number density of galaxies ($\eta_k$), light-weighted stellar age ($\log (\mathrm{Age})$), and effective radius ($R_e$). 
  Error bars denote the $1\sigma$ uncertainties estimated via bootstrap resampling. 
  Statistical significance levels based on the full dataset are indicated above the bars: no asterisks for $p \ge 0.05$ (not significant), a single asterisk ($*$) for $p < 0.05$, and double asterisks ($**$) for $p < 0.001$. 
  Notably, stellar age and the H\,\textsc{i} deficiency parameter maintain statistically significant correlations with the metallicity gradient even after fixing $M_\star$, whereas morphological and structural parameters show negligible and statistically insignificant partial correlations.}
  \label{fig:pcc}
\end{figure*}

Given the important caveats mentioned above regarding the apparent broad consistency between our observations and 
the simulations from \citet{Mercado2021}, we further investigate whether the observed age--$\nabla_r [Z/H]$ relation arises from in-situ star formation or is driven by environmental effects.

To find the galaxy properties relevant to stellar metallicity gradients, we calculate Spearman (partial) correlation coefficients between $\nabla_r [Z/H]$ and a suite of galaxy properties. These include morphological T-type, central stellar mass surface density ($\Sigma_\star$), effective radius ($R_e$), local galaxy number density ($\eta_k$), and the H\,\textsc{i} deficiency parameter ($Def_{\mathrm{HI}}$). We compute the coefficients both with and without controlling for galaxy stellar mass. Figure \ref{fig:pcc} presents the correlation coefficients, which we use in the following subsections to discuss the relevance of various galaxy properties to stellar metallicity gradients.

\subsection{Minimal Dependence on Galaxy Stellar Mass and Morphologies}
\label{sec:mass_ttype}

\begin{figure*}
  \centering
  \includegraphics[width=1\textwidth]{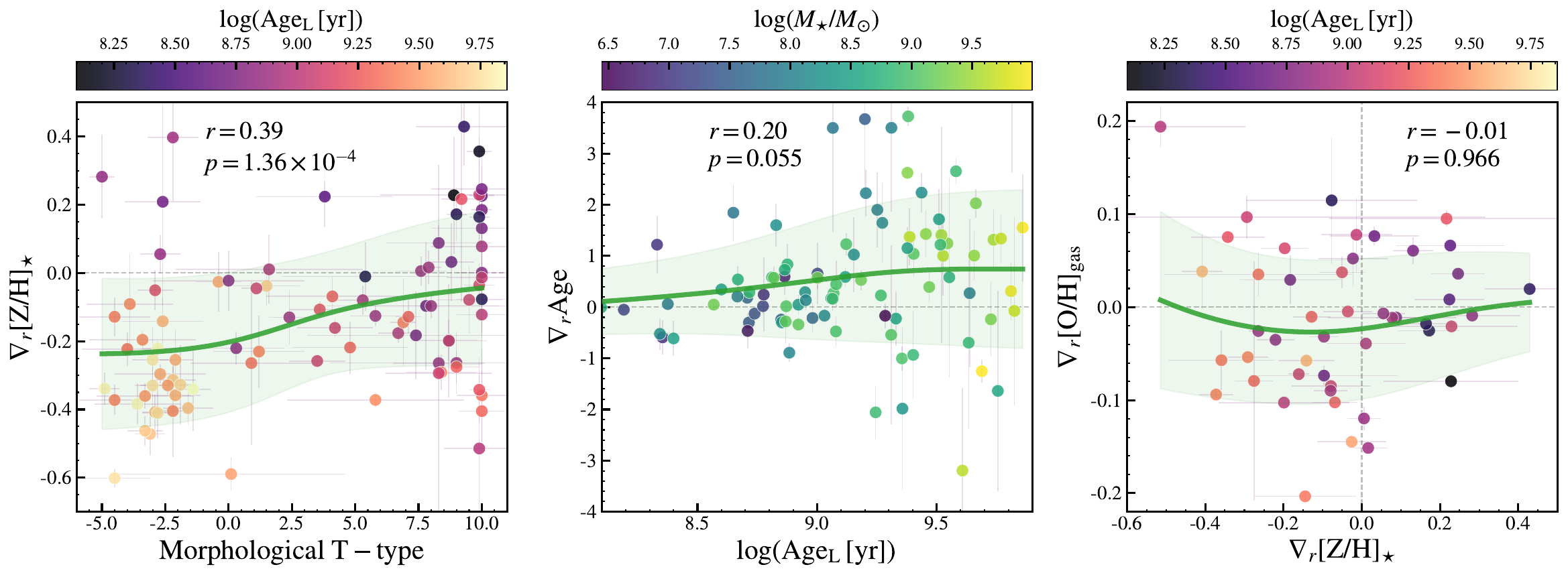}
  \caption{\textbf{Stellar population and gas-phase metallicity gradients across different galaxy scaling relations.} 
  \textit{Left:} Stellar metallicity gradients ($\nabla_r [Z/H]_\star$) as a function of morphological T-Type. Data points are color-coded by light-weighted stellar age ($\log (\mathrm{Age}_\mathrm{L} / \mathrm{yr})$).
  \textit{Middle:} Stellar age gradients ($\nabla_r \mathrm{Age}$) as a function of light-weighted stellar age. Data points are color-coded by stellar mass ($\log (M_\star/M_\odot)$).
  \textit{Right:} Comparison between gas-phase metallicity gradients ($\nabla_r [\mathrm{O/H}]_{\mathrm{gas}}$) and stellar metallicity gradients ($\nabla_r [Z/H]_\star$). Data points are color-coded by light-weighted stellar age ($\log (\mathrm{Age}_\mathrm{L} / \mathrm{yr})$).
  In each panel, the solid green line and its surrounding shaded region denote the global trend and the associated uncertainty, determined via a running median with a Gaussian kernel. 
  The Spearman correlation coefficient ($r$) and the associated $p$-value are  
  The horizontal and vertical lines mark a flat gradient. For more details, see Section \ref{sec:mass_ttype} for the \textit{left} panel, and Section \ref{sec:in-situ} for the \textit{middle} and \textit{right} panels.}
  \label{fig:in-situ}
\end{figure*}

Theoretical study by \cite{Schroyen2011} suggests a ``centrifugal barrier'' governing metallicity gradients: higher specific angular momentum yields more extended star formation and shallower metallicity gradients. 
This aligns with the NIHAO simulations of ultra-diffuse galaxies \citep{CardonaBarrero2023}, which predict nearly age-indpendent shallow metallicity gradients for rotation-supported galaxies whilst a significant age–metallicity gradient relation for dispersion-dominated galaxies. The hatch-shaded region in the right panel of Figure \ref{fig:azgr} represents the parameter space populated by the NIHAO simulations. This dichotomy behavior may partly explain the significant age--metallicity gradient relation found by \cite{Mercado2021}, whose simulated sample is composed almost entirely of dispersion-dominated isolated dwarfs.

Consistent with these theoretical expectations, previous observational studies based on small dwarf samples in dense environment indeed found that dwarf galaxies with flatter stellar metallicity gradients tend to have disky structure and fast rotation \citep[e.g.,][]{Koleva2009}. Since specific angular momentum correlates with stellar mass, and its scatter links to morphological type \citep{Cortese2016, Greene2018} and central surface brightness \citep{Elson2024}, these structural properties should correlate with metallicity gradients if angular momentum is indeed relevant.

Because the SNR and spatial coverage of our MUSE observations are generally insufficient for a robust measurement of stellar angular momentum \citep[see][]{Posti2018}, we instead use morphological T-type, central surface brightness ($\Sigma_{\mathrm{cen}}$), and effective radius ($R_e$) as proxies for kinematic structure. We explore their correlations with metallicity gradients, both before and after controlling for stellar mass (Figure \ref{fig:pcc}). The T-type exhibits a moderate correlation with $\nabla_r [Z/H]$ of $|r| \approx 0.4$ (also see the left panel of Figure \ref{fig:in-situ}), decreasing to $|r| \approx 0.3$ after controlling for stellar mass. The correlations with $R_e$ and $\Sigma_{\mathrm{cen}}$ are negligibly weak ($|r| < 0.2$).

The trend that later morphological types exhibit shallower average metallicity gradients (with substantial scatter) qualitatively follows the theoretical expectation above. However, the dependence on T-Type weakens after controlling for stellar mass and disappears when controlling for stellar ages (partial correlation $|r| <$ 0.02). In contrast, the correlation between the metallicity gradient and light-weighted stellar age remains strong after controlling for stellar mass ($|r| \gtrsim 0.7$), suggesting that stellar age is a more fundamental property determining the metallicity gradient.

\subsection{Limited Impact of Inside‑Out Growth}
\label{sec:in-situ}

The ``inside-out'' formation model traditionally predicts negative metallicity gradients, as central regions enrich earlier and more efficiently than the outskirts \citep[e.g.,][]{Matteucci1989, Mo1998, Boissier1999, Chiappini2001}. Recent simulations further suggest that gas-phase gradients should steepen over time under the inside-out framework \citep{Graf2025}. If subsequent generations of stars simply inherit local gas metallicities, one would expect steeper metallicity gradients in younger populations. This scenario is consistent with observations of the Milky Way, where younger stellar populations indeed exhibit more negative gradients \citep[e.g.,][]{Anders2023}.
However, our sample displays the opposite trend: galaxies with younger average ages possess flatter (less negative) or even positive gradients (Figure \ref{fig:azgr}). That said, we note that the evolution of the metallicity gradients in the framework of inside-out growth remains highly uncertain, partly due to our limited knowledge of cosmic gas accretion \citep{Sharda2021} and feedback/turbulence-driven chemical mixing\citep[e.g.,][]{Hemler2021}.

We further  examine the age gradients ($\nabla_r \mathrm{Age}$). As shown in the middle panel of Figure \ref{fig:in-situ}, the $\nabla_r \mathrm{Age}$ distribution is skewed towards positive values (62\% of the sample) and older galaxies tend to have a more positive median $\nabla_r \mathrm{Age}$ (though with marginal statistical significance). This is inconsistent with inside-out formation histories. A positive age gradient implies that star formation has been recently enhanced in the central regions, that star formation in the outskirts subsided earlier, or that long-term radial migration has redistributed older populations (see Section \ref{sec:migration}). We revisit these possibilities by incorporating environmental effects in Section \ref{sec:envir}. 

The marginal trend of older galaxies having more positive median $\nabla_r \mathrm{Age}$ is consistent with that found in the FIRE simulations by \citet{Graus2019}, though the correlation strength is much stronger in the simulations. As in the FIRE-2 simulations (Section \ref{sec:migration}), this age-dependent trend in the FIRE simulations is mainly caused by feedback-driven stellar migration, so the same caveats for FIRE-2 mentioned in Sections \ref{sec:migration} and \ref{sec:mass_ttype} apply.

The right panel of Figure \ref{fig:in-situ} presents the relation between our stellar metallicity gradients and the gas-phase oxygen abundance gradients from \citet{Li2025}, for the 47 galaxies in common. Given that the stellar metallicity and gas-phase oxygen abundance probe stellar populations formed over different timescales \citep[e.g.,][]{Zhuang2024}, we do not expect a tight relation between the two. That said, a sheer lack of correlation between their gradients, as demonstrated in Figure \ref{fig:in-situ}, points to a diversity of spatial and temporal variations in gas inflows and outflows \citep[e.g.,][]{Lian2018}, and/or significant stellar redistribution.

\subsection{Environmental Quenching and Outside-In Gas Stripping}
\label{sec:envir}

\begin{figure*}
  \centering
  \includegraphics[width=1\textwidth]{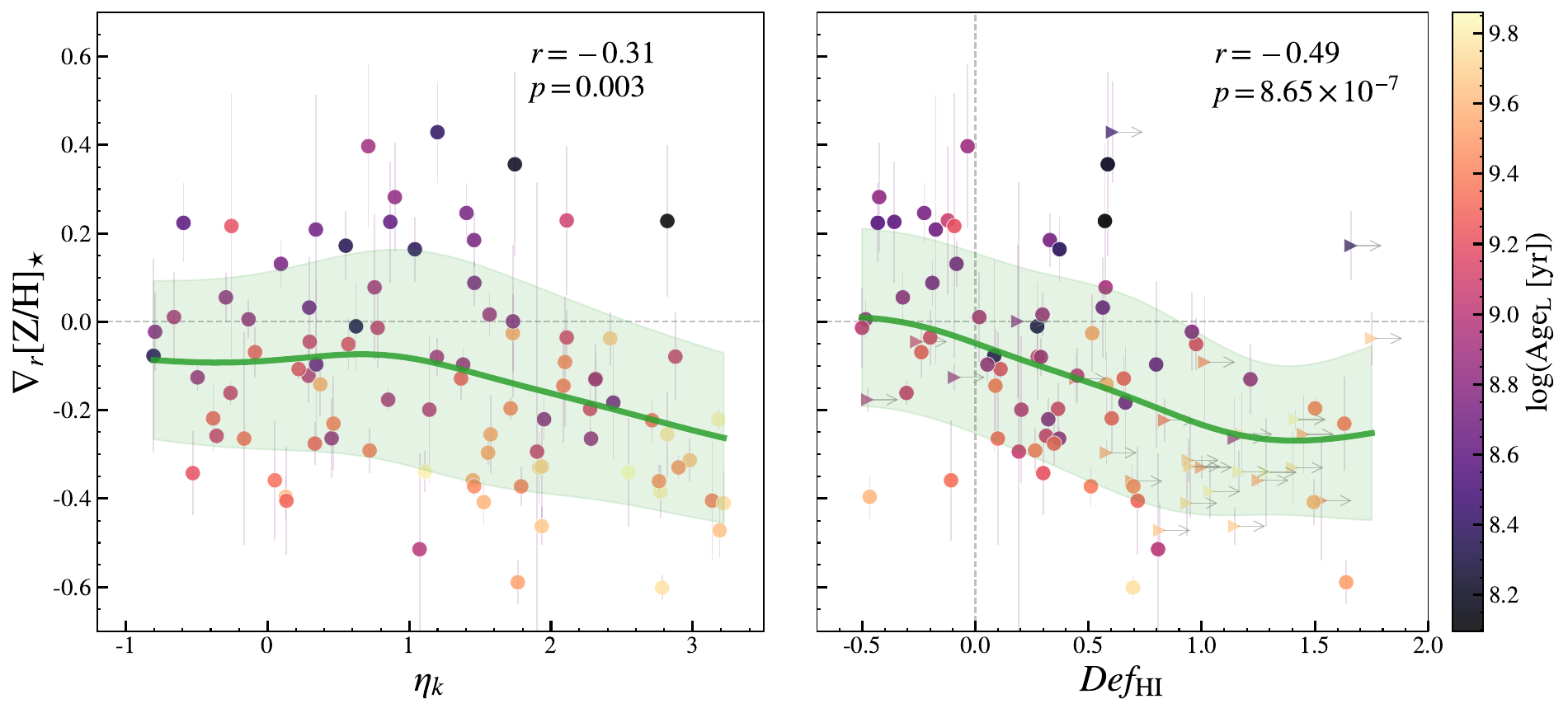}
  \caption{
  \textbf{Environmental dependence of metallicity gradients ($\nabla_r [Z/H]$). }
  \textit{Left:} Metallicity gradients as a function of the local projected density ($\eta_k$). 
  \textit{Right:} Metallicity gradients as a function of the H\,\textsc{i} deficiency ($Def_{\mathrm{HI}}$). 
  All galaxies are color-coded by their light-weighted stellar age ($\log \mathrm{Age} [\mathrm{yr}]$). 
  In the right panel, right-pointing triangles with arrows represent galaxies with H\,\textsc{i} upper limits (lower limits for $Def_{\mathrm{HI}}$). 
  In each panel, the solid green line and the surrounding shaded band represent the overall trend and its uncertainty, derived from a Gaussian running median. 
  The Spearman correlation coefficient ($r$) and the $p$-value are displayed in each panel. 
  Gray dashed horizontal and vertical lines mark the zero-point for each parameter.
  }
  \label{fig:envir}
\end{figure*}

The prevalence of positive age gradients in dwarf galaxies (Section \ref{sec:in-situ}) may result from an ``outside‑in'' shrinking of the star‑forming disk, first noted by \cite{Zhang2012}. Unlike massive systems, dwarfs are more susceptible to environmental quenching, which typically proceeds from the outskirts inward. This process truncates metal enrichment from the outside in, thereby steepening the metallicity gradients.

As shown in the left panel of Figure \ref{fig:envir}, the metallicity gradient exhibits a weak ($r = -0.31$) but statistically significant ($p = 0.003$) correlation with the local projected density ($\eta_k$). However, the correlation strength becomes even weaker if controlling for the stellar mass (Figure \ref{fig:pcc}). The other indicator of environmental effect, H\,\textsc{i} deficiency parameter $Def_{\mathrm{HI}}$, shows a moderate, highly significant negative correlation ($r = -0.49$, $p < 10^{-6}$; right panel of Figure \ref{fig:envir}). This relationship remains robust ($r = -0.43$, $p \sim 10^{-5}$) when excluding lower-limit $Def_{\mathrm{HI}}$ measurements. $Def_{\mathrm{HI}}$ shows a correlation strength with $\nabla_r [Z/H]$ that is second only to Age$_{\rm L}$, even after controlling for stellar mass (Figure \ref{fig:pcc}).

This distinction of the two environmental indicators is physically instructive: while $\eta_k$ is subject to foreground and background contamination and is at most an indicator of the {\it current} local environment, $Def_{\mathrm{HI}}$ acts as a ``fossil record'' of cumulative environmental stripping. Nevertheless, gas stripping operates over diverse timescales—ranging from several hundred Myr to several Gyr—depending on the orbital parameters and host properties. Consequently, the stellar population (which reflects the long-term chemical history) is unlikely to respond to gas removal and the ensuing star formation quenching in perfect lockstep. This temporal mismatch can explain the substantial scatter observed in the $\nabla_r [Z/H]$--$Def_{\mathrm{HI}}$ relation. Despite this scatter, the persistence of the correlation indicates that the physical removal of gas is a tangible driver in shaping the chemical structure of dwarf galaxies.

\section{Summary}
\label{sec:summary}

We present a comprehensive analysis of stellar metallicity radial gradients ($\nabla_r [Z/H]$) for a sample of 90 nearby low-mass galaxies, utilizing high-quality VLT/MUSE integral field spectroscopy. Spanning a stellar mass range of $10^{6.5}$ to $10^{10} M_\odot$ (median $\sim 10^{8.5} M_\odot$), our sample significantly extends the parameter space of existing IFU surveys, which are biased toward more massive systems or lack the sensitivity required for stellar population analysis in the dwarf regime.

Our primary finding is a robust negative correlation between metallicity gradients and light-weighted stellar age (Spearman $|r| \gtrsim 0.7$): older dwarf galaxies have steeper (more negative) gradients, while younger systems show flatter or positive profiles. This relation holds regardless of stellar mass, effective radius, central surface brightness, or large-scale environment (group/field). The correlation strength peaks in the intermediate-mass regime ($8.2 \lesssim \log M_\star/M_\odot \lesssim 9.0$), where the age--$\nabla_r [Z/H]$ relation exhibits its steepest slope.

The slope of the relation varies gradually with the outermost radius limit used for gradient fitting, reflecting that the metallicity profile generally flattens with radius. For the radial range ($\lesssim$ 2$R_{\rm e}$) covered by the observations, the slope of the mass-weighted age--$\nabla_r [Z/H]$ relation is close to that in the FIRE-2 simulations \citep{Mercado2021}, suggesting that stellar radial migration induced by feedback-driven gravitational potential fluctuations may play a fundamental role in dwarf galaxy evolution. However, this agreement is likely coincidental and the impact of stellar migration is likely overestimated: FIRE-2's feedback and chemical mixing are overly efficient, and the simulated galaxies in \citet{Mercado2021} are nearly all dispersion dominated, both inconsistent with observations.

The prevalence of positive radial gradients in stellar ages among our galaxies indicates that most of them have experienced outside-in quenching rather than inside-out growth. We find that the H\,\textsc{i} deficiency parameter ($Def_{\mathrm{HI}}$), an indicator of past environmental stripping, shows a moderate yet highly significant correlation with $\nabla_r [Z/H]$, second only to stellar age in strength. The correlation is in the sense that galaxies which are more H\,\textsc{i} deficient tend to have more negative $\nabla_r [Z/H]$, strongly indicating that environment-driven outside-in quenching and the ensuing truncation of metal enrichment play an important role in shaping the stellar metallicity distribution. 

Our analysis suggests that the chemical evolution of dwarf galaxies likely arises from a synergy between feedback‑driven dynamical heating and external environmental processing, though only the latter has robust observational support. Numerical simulations incorporating more realistic treatments of baryonic feedback are necessary to place compelling constraints on the importance of long‑term stellar radial migration. Obtaining high‑quality stellar spectra of dwarf galaxies at high redshift remains challenging, even with the James Webb Space Telescope (JWST). However, such observations would allow us to witness the gradual buildup of stellar metallicity distributions, offering strong clues to the chemical and structural evolution history of dwarf galaxies.
%TC:ignore
\begin{acknowledgements} 
We acknowledge support from National Key Research and Development Program of China (grant No. 2023YFA1608100), and from the NSFC grant (Nos. 12122303, 11973039, 11421303, 11973038, 12233008). This work is also supported by the China Manned Space Project (Nos.CMS-CSST-2021-B02, CMS-CSST-2021-A07). We acknowledge support from the CAS Pioneer Hundred Talents Program, the Strategic Priority Research Program of Chinese Academy of Sciences (Grant No. XDB 41000000) and the Cyrus Chun Ying Tang Foundations. GC acknowledges financial support provided by the Spanish Ministerio de Ciencia, Innovación y Universidades (MICIU) through the project PID2023-153342NB-I00 / 10.13039/501100011033.
\end{acknowledgements}

\appendix

\section{Dependence of Age--Metallicity Gradient Relations on Radial Boundaries}
\label{sec:appendix_radial_tests}

\begin{figure*}
  \centering
  \includegraphics[width=\textwidth]{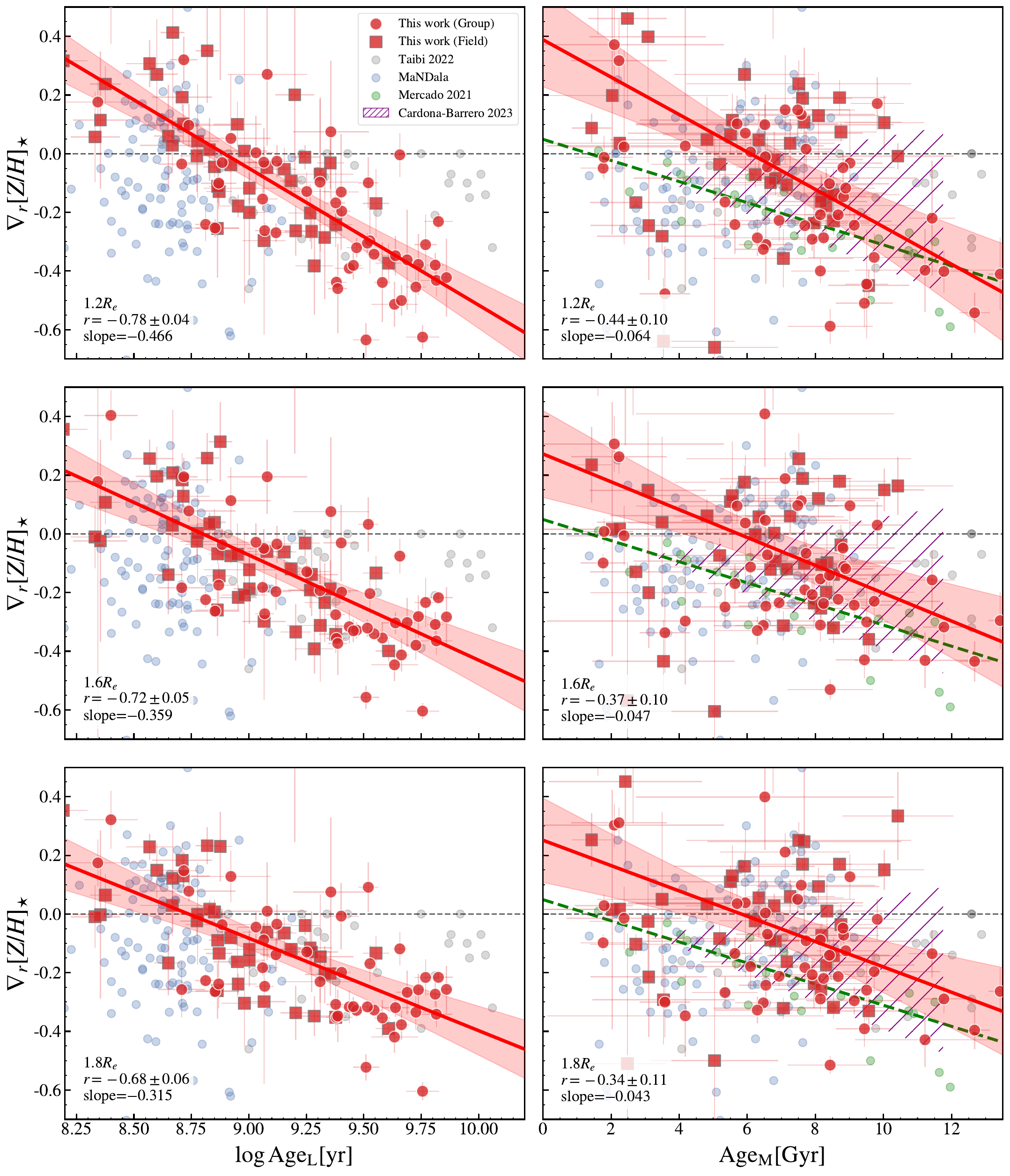}
  \caption{Same as the main plots in Figure \ref{fig:azgr}, but with the metallicity gradients derived using different maximum fitting radii. The maximum fitting radius (in units of $R_{\rm e}$), the age-metallicity gradient correlation coefficient and the best-fit slope are listed in the bottom-left corner of each panel.}
  \label{fig:differ_r_appendix}
\end{figure*}

Figure \ref{fig:differ_r_appendix} presents the stellar age--metallicity gradient relation derived using three different outer radial boundaries for the metallicity gradient fitting. The left panels plot the metallicity gradient as a function of light-weighted ages ($\log \mathrm{Age}_{\mathrm{L}}$), while the right panels plot the metallicity gradient as a function of mass-weighted ages ($\mathrm{Age}_{\mathrm{M}}$). From the top to bottom, different rows show results for increasing outermost radii used in the metallicity gradient fitting: $1.2\,R_e$, $1.6\,R_e$, and $1.8\,R_e$.

\section{Main properties of the galaxy sample}
In this section we present a table with general information and derived properties for all the galaxies in the sample. The meaning of each column is indicated in the notes below.
\begin{longrotatetable}
\begin{deluxetable*}{lllllllllllrrr}
\tablecaption{Main properties of the galaxy sample. \label{tab:property}}
\tablewidth{0pt}
\tablehead{
\colhead{Galaxy} & \colhead{R.A.} & \colhead{Dec.} & \colhead{$z$} & \colhead{$D$} & \colhead{$R_e$} & \colhead{$i$} & \colhead{$\eta_k$} & \colhead{$Def_{\rm HI}$} & \colhead{$\log M_{\star}$} & \colhead{$\log \rm Age$} & \colhead{$\nabla_{r} \rm Age$}& \colhead{[Z/H]} & \colhead{$\nabla_{r} \rm [Z/H]$} \\
\colhead{-} & \colhead{deg} & \colhead{deg} & \colhead{-} & \colhead{Mpc} & \colhead{kpc} & \colhead{deg} & \colhead{Mpc$^{-2}$} & \colhead{dex} & \colhead{$M_{\odot}$} & \colhead{yr} & \colhead{dex/$R_e$}& \colhead{dex} & \colhead{dex/$R_e$} \\
\colhead{(1)} & \colhead{(2)} & \colhead{(3)} & \colhead{(4)} & \colhead{(5)} & \colhead{(6)} & \colhead{(7)} & \colhead{(8)} & \colhead{(9)} & \colhead{(10)} & \colhead{(11)} & \colhead{(12)}& \colhead{(13)} & \colhead{(14)}
}
\startdata
AGC191702 & 137.15226 & 5.29078 & 0.00199 & 12.21 & 0.56 & 18.5 & -0.80 & 0.08 & 6.96 & 8.35 & $-0.59 \pm 0.33$ & -1.39 & $-0.08 \pm 0.22$ \\
AGC193816 & 140.36375 & 7.36639 & 0.00463 & 21.47 & 1.18 & 0.0 & 0.05 & -0.11 & 8.48 & 9.33 & $-0.23 \pm 0.53$ & -1.52 & $-0.36 \pm 0.14$ \\
CGCG007-025 & 146.00780 & -0.64227 & 0.00481 & 23.05 & 1.05 & 62.2 & 1.20 & 0.61 & 8.46 & 8.40 & $-0.61 \pm 0.34$ & -1.55 & $0.43 \pm 0.11$ \\
CGCG035-007 & 143.68627 & 6.42563 & 0.00182 & 4.92 & 0.25 & 47.4 & 0.09 & -0.08 & 7.37 & 8.71 & $-0.30 \pm 0.23$ & -0.82 & $0.13 \pm 0.05$ \\
ESO115-021 & 39.41943 & -61.35170 & 0.00170 & 4.99 & 1.06 & 74.0 & -0.13 & -0.48 & 7.58 & 8.85 & $-0.24 \pm 0.07$ & -1.14 & $0.01 \pm 0.04$ \\
ESO157-49 & 69.90404 & -53.01225 & 0.00560 & 19.32 & 1.92 & 79.5 & -0.38 & 0.60 & 8.90 & 9.27 & $0.23 \pm 0.06$ & -0.70 & $-0.22 \pm 0.08$ \\
ESO158-003 & 71.57227 & -57.34378 & 0.00400 & 10.20 & 1.48 & 30.0 & 0.55 & 1.66 & 8.14 & 8.34 & $-0.52 \pm 0.31$ & -0.50 & $0.17 \pm 0.08$ \\
ESO162-017 & 108.97637 & -57.34381 & 0.00366 & 11.86 & 1.45 & 73.2 & -0.36 & 0.31 & 9.02 & 9.07 & $0.27 \pm 0.09$ & -0.81 & $-0.26 \pm 0.02$ \\
ESO184-82 & 293.76842 & -52.84389 & 0.00869 & 28.58 & 1.79 & 47.2 & -0.09 & -0.24 & 8.89 & 9.24 & $-2.06 \pm 0.51$ & -0.87 & $-0.07 \pm 0.06$ \\
ESO320-14 & 174.47182 & -39.22034 & 0.00218 & 6.00 & 0.70 & 56.4 & 0.29 & 0.45 & 7.14 & 9.00 & $0.65 \pm 0.60$ & -1.30 & $-0.12 \pm 0.06$ \\
ESO349-031 & 2.05567 & -34.57833 & 0.00076 & 3.21 & 0.42 & 0.0 & 1.46 & 0.33 & 6.58 & 8.71 & $-0.47 \pm 0.34$ & -1.39 & $0.18 \pm 0.06$ \\
ESO379-024 & 181.23612 & -35.74303 & 0.00210 & 5.47 & 0.53 & 59.1 & 0.62 & 0.27 & 6.98 & 8.33 & $1.22 \pm 0.56$ & -1.40 & $-0.01 \pm 0.10$ \\
ESO469-15 & 347.23192 & -30.85838 & 0.00546 & 25.46 & 1.41 & 75.4 & 0.22 & 0.11 & 9.23 & 9.18 & $0.53 \pm 0.19$ & -0.62 & $-0.11 \pm 0.02$ \\
ESO483-013 & 63.17129 & -23.15887 & 0.00270 & 10.68 & 1.36 & 46.0 & 0.37 & 0.58 & 8.67 & 9.55 & $0.58 \pm 1.95$ & -1.04 & $-0.14 \pm 0.07$ \\
ESO486-021 & 75.83203 & -25.42293 & 0.00290 & 9.11 & 0.64 & 30.0 & -0.26 & -0.30 & 8.10 & 9.06 & $0.17 \pm 0.04$ & -1.22 & $-0.16 \pm 0.03$ \\
ESO495-021 & 129.06300 & -26.40949 & 0.00291 & 8.23 & 0.32 & 53.3 & 0.34 & -0.17 & 8.65 & 8.67 & $0.54 \pm 0.27$ & -0.63 & $0.21 \pm 0.30$ \\
FCC047 & 51.63415 & -35.71351 & 0.00473 & 18.30 & 1.55 & 54.0 & 1.92 & 1.40 & 9.54 & 10.10 & $1.32 \pm 0.58$ & -0.70 & $-0.33 \pm 0.08$ \\
FCC090 & 52.78442 & -36.29014 & 0.00605 & 19.23 & 1.10 & 43.0 & 1.79 & 0.51 & 9.00 & 9.38 & $3.73 \pm 0.19$ & -0.72 & $-0.37 \pm 0.04$ \\
FCC113 & 53.27854 & -34.80811 & 0.00463 & 16.14 & 1.60 & 39.0 & 2.28 & 0.37 & 8.45 & 9.12 & $0.59 \pm 0.19$ & -0.87 & $-0.20 \pm 0.06$ \\
FCC119 & 53.39101 & -33.57332 & 0.00458 & 20.14 & 1.45 & 32.5 & 1.53 & 1.50 & 8.88 & 9.58 & $2.65 \pm 0.25$ & -0.61 & $-0.41 \pm 0.05$ \\
FCC143 & 53.74671 & -35.17111 & 0.00445 & 17.54 & 0.71 & 44.2 & 2.77 & 1.04 & 9.60 & 9.82 & $4.10 \pm 1.46$ & -0.41 & $-0.38 \pm 0.06$ \\
FCC148 & 53.82003 & -35.26563 & 0.00259 & 18.08 & 1.49 & 68.4 & 3.14 & 1.53 & 9.69 & 9.39 & $1.37 \pm 0.55$ & -0.15 & $-0.40 \pm 0.13$ \\
FCC182 & 54.22626 & -35.37466 & 0.00565 & 19.41 & 0.94 & 18.6 & 3.22 & 0.94 & 9.30 & 9.73 & $-0.24 \pm 0.72$ & -0.46 & $-0.41 \pm 0.07$ \\
FCC190 & 54.28735 & -35.19500 & 0.00598 & 18.86 & 1.37 & 39.8 & 3.18 & 1.41 & 9.72 & 9.82 & $-0.08 \pm 1.85$ & -0.37 & $-0.22 \pm 0.02$ \\
FCC202 & 54.52735 & -35.43985 & 0.00269 & 19.02 & 1.04 & 52.5 & 3.19 & 0.82 & 8.82 & 9.63 & $-0.70 \pm 1.67$ & -0.72 & $-0.47 \pm 0.06$ \\
FCC207 & 54.58029 & -35.12908 & 0.00474 & 17.50 & 0.76 & 40.4 & 2.78 & 0.70 & 8.54 & 9.76 & $-1.64 \pm 1.96$ & -1.45 & $-0.60 \pm 0.03$ \\
FCC211 & 54.58953 & -35.25973 & 0.00775 & 21.80 & 0.76 & 50.1 & 2.32 & 0.44 & 8.31 & 9.36 & $-1.98 \pm 1.58$ & -0.97 & $-0.13 \pm 0.06$ \\
FCC222 & 54.80542 & -35.37144 & 0.00267 & 8.02 & 0.68 & 30.7 & 2.98 & 0.94 & 7.89 & 9.64 & $0.27 \pm 1.45$ & -0.80 & $-0.31 \pm 0.05$ \\
FCC249 & 55.17544 & -37.51075 & 0.00527 & 23.54 & 0.79 & 29.6 & 1.11 & 1.16 & 9.90 & 9.81 & $0.31 \pm 0.57$ & -0.55 & $-0.34 \pm 0.05$ \\
FCC255 & 55.26504 & -33.77931 & 0.00436 & 19.77 & 1.36 & 64.9 & 1.45 & 1.25 & 9.33 & 9.54 & $1.25 \pm 0.16$ & -0.42 & $-0.36 \pm 0.01$ \\
FCC263 & 55.38583 & -34.88833 & 0.00575 & 12.59 & 1.06 & 65.0 & 1.95 & 0.32 & 8.71 & 8.81 & $0.58 \pm 0.10$ & -1.01 & $-0.22 \pm 0.05$ \\
FCC277 & 55.59471 & -35.15400 & 0.00547 & 20.51 & 1.25 & 55.1 & 2.82 & 1.18 & 9.73 & 9.77 & $1.34 \pm 0.49$ & -0.52 & $-0.25 \pm 0.10$ \\
FCC285 & 55.75914 & -36.27337 & 0.00296 & 9.16 & 1.67 & 53.8 & 1.57 & 0.30 & 8.08 & 8.88 & $-0.89 \pm 0.18$ & -0.86 & $0.02 \pm 0.05$ \\
FCC301 & 56.26483 & -35.97261 & 0.00341 & 16.29 & 0.73 & 45.5 & 1.93 & 1.15 & 9.30 & 9.66 & $2.03 \pm 0.28$ & -0.47 & $-0.46 \pm 0.04$ \\
FCC306 & 56.43916 & -36.34653 & 0.00296 & 9.74 & 0.39 & 45.7 & 1.46 & -0.19 & 7.25 & 8.74 & $-0.13 \pm 0.03$ & -1.24 & $0.09 \pm 0.05$ \\
FCC308 & 56.47854 & -36.35697 & 0.00500 & 8.68 & 1.47 & 72.4 & 2.28 & 0.37 & 8.31 & 8.85 & $-0.30 \pm 0.08$ & -0.57 & $-0.26 \pm 0.02$ \\
FCC332 & 57.45425 & -35.94558 & 0.00442 & 15.49 & 1.04 & 45.5 & 1.77 & 1.64 & 8.39 & 9.51 & $1.72 \pm 0.59$ & -0.83 & $-0.59 \pm 0.05$ \\
FCC335 & 57.65304 & -35.90933 & 0.00477 & 19.23 & 1.44 & 55.8 & 1.72 & 1.50 & 9.08 & 9.40 & $1.04 \pm 0.44$ & -0.57 & $-0.20 \pm 0.03$ \\
IC0217 & 34.04298 & -11.92733 & 0.00630 & 24.13 & 3.72 & 86.7 & -0.49 & -0.09 & 9.05 & 8.87 & $0.02 \pm 0.05$ & -0.68 & $-0.13 \pm 0.03$ \\
IC0719 & 175.07711 & 9.01000 & 0.00611 & 23.00 & 1.51 & 78.9 & 0.13 & -0.47 & 9.63 & 9.61 & $-3.19 \pm 1.61$ & -0.29 & $-0.40 \pm 0.05$ \\
IC0745 & 178.55119 & 0.13672 & 0.00381 & 18.90 & 0.69 & 23.6 & 0.71 & -0.03 & 8.76 & 8.88 & $0.83 \pm 0.40$ & -0.77 & $0.40 \pm 0.19$ \\
IC1574 & 10.76486 & -22.24313 & 0.00120 & 4.92 & 0.98 & 53.0 & 1.75 & 0.58 & 7.10 & 8.19 & $-0.05 \pm 0.05$ & -0.94 & $0.36 \pm 0.21$ \\
IC1959 & 53.30246 & -50.41425 & 0.00213 & 6.85 & 1.27 & 79.5 & 0.72 & 0.26 & 8.45 & 9.38 & $1.15 \pm 0.41$ & -1.23 & $-0.29 \pm 0.05$ \\
IC2828 & 171.79559 & 8.73105 & 0.00345 & 14.20 & 0.70 & 64.9 & 0.76 & 0.58 & 8.34 & 8.92 & $0.05 \pm 0.19$ & -1.17 & $0.08 \pm 0.16$ \\
IC3392 & 187.18043 & 14.99956 & 0.00557 & 11.69 & 1.77 & 67.8 & 2.31 & 1.22 & 9.15 & 8.87 & $-0.28 \pm 0.43$ & -0.26 & $-0.13 \pm 0.08$ \\
IC3476 & 188.17452 & 14.05044 & -0.00053 & 13.80 & 1.97 & 47.8 & 2.88 & 0.28 & 8.92 & 9.06 & $0.16 \pm 0.24$ & -0.59 & $-0.08 \pm 0.10$ \\
IC4247 & 201.68539 & -30.36280 & 0.00140 & 4.97 & 0.47 & 53.0 & 2.44 & 0.66 & 7.55 & 8.71 & $0.18 \pm 0.06$ & -1.25 & $-0.18 \pm 0.13$ \\
IC4870 & 294.40667 & -65.81183 & 0.00292 & 8.51 & 1.04 & 58.6 & 0.87 & -0.36 & 8.37 & 8.60 & $0.34 \pm 0.06$ & -1.00 & $0.23 \pm 0.14$ \\
KKH086 & 208.63999 & 4.24398 & 0.00100 & 2.60 & 0.37 & 61.0 & 0.13 & 0.72 & 6.45 & 9.29 & $-0.17 \pm 0.24$ & -0.40 & $-0.40 \pm 0.12$ \\
LEDA683190 & 173.29542 & -32.96244 & 0.00236 & 5.60 & 0.35 & 40.2 & 0.45 & 1.14 & 6.75 & 8.86 & $0.60 \pm 0.35$ & -1.39 & $-0.26 \pm 0.09$ \\
MCG-03-34-002 & 196.98617 & -16.68920 & 0.00310 & 7.90 & 0.69 & 52.0 & 0.34 & 0.80 & 7.84 & 8.65 & $1.84 \pm 0.55$ & -0.94 & $-0.10 \pm 0.19$ \\
NGC0059 & 3.85488 & -21.44450 & 0.00130 & 5.30 & 0.74 & 47.0 & 0.57 & 0.98 & 8.14 & 9.15 & $1.03 \pm 0.74$ & -0.82 & $-0.05 \pm 0.06$ \\
NGC853 & 32.92161 & -9.30599 & 0.00501 & 21.00 & 2.24 & 58.5 & 1.14 & 0.20 & 9.25 & 9.00 & $0.57 \pm 0.21$ & -0.44 & $-0.20 \pm 0.17$ \\
NGC1311 & 50.02900 & -52.18553 & 0.00190 & 5.45 & 0.87 & 62.0 & 0.29 & 0.56 & 7.91 & 8.67 & $0.21 \pm 0.14$ & -0.97 & $0.03 \pm 0.11$ \\
NGC1705 & 73.55625 & -53.36106 & 0.00211 & 5.10 & 0.43 & 45.6 & -0.29 & -0.32 & 8.23 & 8.83 & $1.60 \pm 0.42$ & -1.23 & $0.06 \pm 0.06$ \\
NGC1796 & 75.67729 & -61.14006 & 0.00338 & 10.60 & 1.31 & 64.0 & 1.20 & 0.29 & 9.14 & 8.92 & $-0.34 \pm 0.13$ & -0.39 & $-0.08 \pm 0.04$ \\
NGC1800 & 76.60717 & -31.95422 & 0.00272 & 8.01 & 0.83 & 63.8 & 1.38 & 0.05 & 8.67 & 8.87 & $0.73 \pm 0.11$ & -0.72 & $-0.10 \pm 0.04$ \\
NGC2915 & 141.54804 & -76.62633 & 0.00156 & 4.29 & 0.32 & 56.4 & -0.66 & 0.02 & 8.20 & 8.95 & $0.29 \pm 0.06$ & -0.97 & $0.01 \pm 0.10$ \\
NGC3125 & 151.63905 & -29.93486 & 0.00371 & 15.00 & 0.82 & 53.6 & 0.90 & -0.42 & 9.17 & 8.82 & $0.57 \pm 0.22$ & -1.11 & $0.28 \pm 0.12$ \\
NGC3274 & 158.07168 & 27.66880 & 0.00176 & 6.50 & 0.64 & 60.7 & 0.85 & -0.48 & 8.03 & 8.95 & $0.13 \pm 0.19$ & -1.05 & $-0.18 \pm 0.03$ \\
NGC3593 & 168.65417 & 12.81767 & 0.00208 & 8.95 & 2.02 & 68.6 & 1.73 & 0.52 & 9.52 & 9.52 & $1.41 \pm 2.11$ & -0.18 & $-0.03 \pm 0.09$ \\
NGC4064 & 181.04647 & 18.44369 & 0.00302 & 9.69 & 1.58 & 76.1 & 0.47 & 1.63 & 9.34 & 9.38 & $2.62 \pm 0.09$ & -0.33 & $-0.23 \pm 0.11$ \\
NGC4150 & 182.64021 & 30.40153 & 0.00069 & 13.06 & 1.06 & 51.6 & 1.57 & 1.45 & 9.62 & 9.53 & $1.00 \pm 0.78$ & -0.27 & $-0.25 \pm 0.09$ \\
NGC4483 & 187.66936 & 9.01567 & 0.00295 & 16.87 & 1.23 & 55.9 & 2.55 & 1.29 & 9.96 & 9.86 & $1.55 \pm 1.05$ & -0.29 & $-0.34 \pm 0.12$ \\
NGC4516 & 188.28140 & 14.57495 & 0.00314 & 15.48 & 1.86 & 65.3 & 2.42 & 1.75 & 9.43 & 9.66 & $1.01 \pm 0.48$ & -0.32 & $-0.04 \pm 0.06$ \\
NGC4809 & 193.71276 & 2.65409 & 0.00294 & 21.90 & 2.98 & 75.0 & 2.11 & -0.12 & 8.98 & 9.08 & $0.45 \pm 0.46$ & -1.31 & $0.23 \pm 0.17$ \\
NGC4810 & 193.713734 & 2.64049 & 0.00294 & 21.90 & 2.56 & 67.6 & 2.11 & -0.20 & 9.00 & 9.12 & $1.23 \pm 0.22$ & -1.19 & $-0.04 \pm 0.06$ \\
NGC4980 & 197.29166 & -28.64183 & 0.00477 & 17.00 & 1.79 & 63.3 & 0.30 & -0.26 & 8.98 & 9.08 & $-0.48 \pm 0.29$ & -1.20 & $-0.05 \pm 0.01$ \\
NGC5253 & 204.98318 & -31.64011 & 0.00136 & 3.55 & 0.70 & 63.9 & 2.82 & 0.57 & 8.65 & 8.10 & $-0.01 \pm 0.03$ & -1.04 & $0.23 \pm 0.17$ \\
NGC5770 & 223.31253 & 3.95975 & 0.00491 & 20.57 & 1.31 & 46.7 & 1.93 & 0.95 & 10.00 & 9.69 & $-1.25 \pm 0.22$ & -0.45 & $-0.33 \pm 0.12$ \\
PGC132213 & 330.78995 & -12.37181 & 0.00275 & 9.31 & 0.32 & 46.5 & -0.79 & 0.96 & 7.13 & 8.77 & $0.02 \pm 0.14$ & -1.87 & $-0.02 \pm 0.09$ \\
PGC166152 & 196.25875 & -40.08278 & 0.00206 & 5.94 & 0.62 & 27.7 & 1.73 & 0.19 & 6.87 & 8.78 & $0.24 \pm 0.73$ & -1.86 & $0.00 \pm 0.17$ \\
UGC685 & 16.84350 & 16.68457 & 0.00050 & 4.70 & 0.83 & 35.0 & -0.25 & -0.09 & 7.51 & 9.20 & $3.67 \pm 1.51$ & -1.14 & $0.22 \pm 0.30$ \\
UGC695 & 16.94350 & 1.06367 & 0.00220 & 10.68 & 1.50 & 32.0 & 1.90 & 0.19 & 7.85 & 9.07 & $3.50 \pm 2.03$ & -1.53 & $-0.29 \pm 0.61$ \\
UGC891 & 20.32915 & 12.41251 & 0.00210 & 11.10 & 1.38 & 52.0 & 0.33 & 0.35 & 8.06 & 9.27 & $1.65 \pm 0.77$ & -1.43 & $-0.28 \pm 0.05$ \\
UGC1056 & 22.19739 & 16.68840 & 0.00200 & 10.57 & 0.52 & 52.6 & 1.04 & 0.37 & 7.83 & 8.38 & $0.06 \pm 0.05$ & -1.10 & $0.16 \pm 0.07$ \\
UGC3755 & 108.46500 & 10.52194 & 0.00105 & 6.67 & 0.79 & 62.1 & -0.53 & 0.30 & 8.02 & 9.20 & $2.23 \pm 0.46$ & -1.36 & $-0.34 \pm 0.09$ \\
UGC5923 & 162.28139 & 6.91742 & 0.00240 & 7.33 & 0.31 & 52.0 & -0.16 & 0.10 & 7.99 & 9.31 & $3.50 \pm 1.64$ & -1.10 & $-0.26 \pm 0.24$ \\
UGC8041 & 193.80273 & 0.11665 & 0.00442 & 14.52 & 3.43 & 40.0 & 2.09 & 0.09 & 9.01 & 9.36 & $-1.00 \pm 0.23$ & -0.67 & $-0.14 \pm 0.13$ \\
UGCA116 & 88.92750 & 3.39222 & 0.00268 & 14.43 & 1.06 & 53.2 & -0.59 & -0.43 & 9.25 & 8.57 & $0.05 \pm 0.11$ & -1.50 & $0.22 \pm 0.09$ \\
UGCA193 & 150.65047 & -6.01371 & 0.00220 & 9.70 & 1.28 & 77.0 & 1.37 & 0.66 & 7.94 & 9.25 & $1.90 \pm 0.73$ & -1.15 & $-0.13 \pm 0.05$ \\
UM461 & 177.88896 & -2.37276 & 0.00347 & 19.54 & 0.63 & 48.4 & 0.78 & -0.50 & 8.06 & 9.03 & $-0.17 \pm 0.12$ & -1.67 & $-0.01 \pm 0.09$ \\
UM462 & 178.15497 & -2.46942 & 0.00347 & 19.18 & 1.07 & 26.7 & 1.40 & -0.23 & 8.82 & 8.72 & $0.29 \pm 0.10$ & -1.50 & $0.25 \pm 0.07$ \\
VCC0170 & 183.98475 & 14.43303 & 0.00466 & 9.12 & 1.28 & 51.6 & 1.46 & 0.70 & 8.19 & 9.44 & $2.24 \pm 0.39$ & -0.67 & $-0.37 \pm 0.02$ \\
VCC0608 & 185.75721 & 15.90558 & 0.00607 & 28.18 & 3.29 & 57.7 & 2.90 & 1.00 & 9.24 & 9.47 & $0.39 \pm 0.30$ & -0.66 & $-0.33 \pm 0.03$ \\
VCC0794 & 186.34000 & 16.42944 & 0.00558 & 26.55 & 3.40 & 63.9 & 2.71 & 0.84 & 9.05 & 9.31 & $-0.54 \pm 1.75$ & -0.90 & $-0.22 \pm 0.05$ \\
VCC0990 & 186.82058 & 16.02447 & 0.00568 & 26.49 & 1.27 & 45.6 & 2.76 & 0.69 & 9.37 & 9.46 & $1.43 \pm 0.47$ & -0.62 & $-0.36 \pm 0.04$ \\
VCC1833 & 190.08199 & 15.93528 & 0.00569 & 16.14 & 0.71 & 46.2 & 1.56 & 0.58 & 8.87 & 9.52 & $1.22 \pm 0.33$ & -0.49 & $-0.30 \pm 0.03$ \\
VCC2019 & 191.33510 & 13.69266 & 0.00607 & 16.98 & 1.51 & 48.0 & 2.10 & 1.01 & 8.84 & 9.40 & $-0.93 \pm 0.85$ & -0.52 & $-0.09 \pm 0.15$ \\
VCC2037 & 191.56375 & 10.20556 & 0.00381 & 9.63 & 1.49 & 66.6 & 1.07 & 0.81 & 7.55 & 8.98 & $-0.21 \pm 0.30$ & -1.51 & $-0.51 \pm 0.22$ \\
\enddata
\tablecomments{
Column (1): Galaxy name.
Column (2)--(3): Right ascension and declination in J2000.
Column (4): Redshift.
Column (5): Distance in Mpc.
Column (6): Effective radius in kpc.
Column (7): Inclination angle in degrees.
Column (8): Local environmental density $\eta_k$.
Column (9): H\,\textsc{i} deficiency parameter.
Column (10): Total stellar mass in $\log M_\odot$.
Column (11)--(12): Light-weighted stellar age and its radial gradient.
Column (13)--(14): Light-weighted stellar metallicity and its radial gradient.
}
\end{deluxetable*}
\end{longrotatetable}

\bibliography{reference}{}
\bibliographystyle{aasjournalv7}
%TC:endignore
\end{document}